\documentstyle[eqsecnum,aps,epsfig]{revtex}

\begin{document}
\draft
\preprint{HEP/123-qed}
\title{Static SU(3) potentials for sources in various representations}
\author{Sedigheh Deldar}
\address{
Washington University, St. Louis, MO, 63130
}
\date{\today}
\maketitle

\begin{abstract}

The potentials and string tensions between static sources in a variety 
of representations (fundamental, 8, 6, 15-antisymmetric, 10, 27 and 
15-symmetric) have been computed by measuring Wilson loops in pure 
gauge $SU(3)$. The simulations have been done primarily on anisotropic 
lattices, using a tadpole improved action improved to $O(a_{s}^4)$. 
A range of lattice 
spacings (0.43 fm, 0.25 fm and 0.11 fm) and volumes ($8^3\times 24$, 
$10^3 \times 24$, $16^3 \times 24$ and $18^3 \times 24$) has been used 
in an attempt to control discretization and finite volume effects.
At intermediate distances, the results show approximate Casimir scaling. 
Finite lattice spacing effects dominate systematic error, and are
particularly large for the representations with the largest string
tensions.
 
\end{abstract}

\maketitle

\section{INTRODUCTION}

One of the main goals of a non-perturbative formulation of gauge theories
is to understand the phenomenon of confinement.
According to the theory of confinement, the potential energy of a pair
of heavy static quarks should increase linearly as a string or tube of
flux is formed between them. For confined representations, one expects
to see a Coulombic plus a linear term for the potential,
$V(R) \simeq -A/R + KR +C $ , where $K$ is the string tension and
$R$ is the quark separation. For screened representations, $V(R)$ should
level off at very large $R$, but one still expects the confined form for
intermediate $R$ with approximate Casimir scaling
\footnote{Here, the Casimir scaling means that the string tension for each
representation is to be roughly proportional to the eigenvalue of
the quadratic Casimir operator in that representation. The Casimir 
scaling regime is expected to exist for intermediate distances, perhaps 
extending from the onset of confinement to the onset of screening 
\cite{Fabe97}.}  of the string tension.
Color screening must occur for adjoint quarks at sufficiently large
separation, but it is very difficult to observe in numerical
simulations, at least for zero temperature. The string tensions in
various representations are basic properties of the pure non-Abelian
gauge theory. To be viable, any theory of confinement (monopole
condensation, center vortices,...) should thus be expected to reproduce
the string tension in all representations.

The formation of a flux
tube and the linear confinement of static quarks in the fundamental
representation has been well established by pure gauge lattice QCD.
The confinement of static sources in higher representations of QCD is 
still a question. Some numerical calculations have been done for SU(2) 
\cite{Bern82} and SU(3) \cite{Fabe88} at non-zero temperature. At 
zero temperature, there have been studies for 8 (adjoint) \cite{Camp86,Mich92},
and 6 and 8 \cite{Mich98}.
Preliminary results of the string tensions of some higher representations
at zero temperature appear in \cite{Deld99,Del299}. Independent of this work, 
G.\ Bali has recently reported some results of the static source
potentials of some higher representations \cite{Bali99,Bali00}.

In this paper, I give results for the potentials
and string tensions of static sources of fundamental and some higher
representations (6, 8, $15_{a}$, 10, 27, $15_{s}$) in pure SU(3).
Most of the simulations have been done on anisotropic lattices with
a tadpole improved action \cite{Morn97}.
The main focus of this paper is the study of potentials at intermediate 
distances.
I show that the potential in each representation at intermediate distances
has a linear behavior and that the string tension is representation dependent 
and roughly proportional to the quadratic Casimir number.

\section{CALCULATIONS}

The string tension may be found by measuring the Wilson loops and looking
for the area law falloff for large $T$, $W(R,t)\simeq \exp^{-V(R)T}$ where
$W(R,T)$ is the Wilson loop as a function of $R$, the
spatial separation of the quarks, and the propagation time $T$, and $V(R)$
is the gauge field energy associated with the static quark-antiquark 
source.  The interquark energy for large separation grows linearly and 
for small separation is expected to be Coulombic, from single gluon 
exchange.

Direct measurement of Wilson loops for higher representations by
multiplying the large matrices is not feasible considering the computer
memory and the running time. One may expand the trace of
Wilson loops for higher representations in terms of Wilson loops, $U$, in
the fundamental representation. The higher representation states are
defined by the tensor product method. Let $W$ be
the higher representation counterpart to fundamental $U$ for 6, 8, 10,
15-symmetric, 15-antisymmetric, or 27, then:
\begin{equation}
6:~tr(W) = 1/2~ [~(trU)^2 + trU^2)~ ],
\label{r6}
\end{equation}
\begin{equation}
8:~tr(W) = trU^\star trU -1,
\label{r8}
\end{equation}
\begin{eqnarray}
10: ~tr(W) = 1/6~[~(trU)^3 + 2(trU^3) + 3trUtrU^2~ ],
\label{r10}
\end{eqnarray}
\begin{eqnarray}
15s:~tr(W) = 1/24~[~ (trU)^4 + 6(trU)^2trU^2  \nonumber \\
              +8trU(trU^3)
              +3(trU^2)^2 + 6trU^4 ~],
\label{r15s}
\end{eqnarray}
\begin{eqnarray}
15a:~tr(W) = 1/2trU^\star~[~(trU)^2+ trU^2]-trU,
\label{r15a}
\end{eqnarray}
\begin{eqnarray}
27:~tr(W)=1/4[trU^2+(trU)^2]~[(tr(U^\star)^2)+(trU^\star)^2~]-trUtrU^\star,
\label{r27}
\end{eqnarray}
where $\star$ indicates complex conjugate.

\section{SIMULATIONS}

Two different improved actions are used for the lattice simulations, one
for isotropic and the other one for anisotropic lattices. Using improved
actions enables one to use coarser lattices and thereby to save computer 
time. 

Calculating the static quark potential by measuring Wilson loops requires
measurements at large enough $T$. Using a coarse isotropic lattice with
fewer temporal sites than a usual fine lattice makes it difficult to
measure quantities like the static quark potential. Anisotropic lattices,
with a temporal lattice spacing smaller than the spatial one, make
possible more measurements in the time direction. Since the signal to
noise ratio of the correlation functions calculated in Monte Carlo
simulations decays exponentially in time, having time slices closer
together enables one to obtain more accurate results. The cost of
simulations would still be less than the cost for fine lattices because
of the coarseness of the spatial lattice spacings. 

\subsection{Isotropic lattice}

The one-loop, tadpole-improved QCD gauge action
of Ref.\ \cite{Dimm95} is used for lattice simulations:
\begin{eqnarray}
S[U]~ = ~ \beta_{pl} \sum_{pl} \frac{1}{3} Re Tr(1-U_{pl}) + 
      \beta_{rt} \sum_{rt} \frac{1}{3} Re Tr(1-U_{rt}) + 
      \beta_{pg} \sum_{pg} \frac{1}{3} Re Tr(1-U_{pg}).
\label{isotropic}
\end{eqnarray}
$U_{pl}$ is the usual $1\times 1$ plaquette operator, $U_{rt}$ is the
$1\times 2$ rectangle operator, and $U_{pg}$ is the six-link parallelogram
operator (path $\mu,\nu,\rho,-\mu,-\nu,-\rho$ where $\mu, \nu, \rho$ are
all different directions).
For this action the plaquette coupling is taken as an
input and the other two couplings are computed using one-loop perturbation
theory \cite{Bali92}:
\begin{eqnarray}
\label{beta-rt}
\beta_{rt}= - \frac{\beta_{pl}}{20u_{0}^2}(1+0.4805 \alpha_{s}), 
\end{eqnarray}
\begin{eqnarray}
\label{beta-pg}
\beta_{pg}= - \frac{\beta_{pl}}{u_{0}^2} 0.03325 \alpha_{s}.
\end{eqnarray}
The mean link $u_{0}$ and the QCD coupling constant $\alpha_{s}$ are as
follows:
\begin{eqnarray}
u_{0}=(\frac{1}{3}Re Tr\langle U_{Pl}\rangle)^{\frac{1}{4}}, 
\end{eqnarray}
\begin{eqnarray}
\alpha_{s}=\frac {\frac{1}{3}Re Tr (1-\langle U_{pl}\rangle)} {3.06839}.
\label{alpha-s}
\end{eqnarray}
The discretization error is of order $O(\alpha_{s}^2 a^2, a^4)$.

Using this action, the fundamental representation static quark potentials
for the following on- and off-axis points are found on a $8^4$ lattice at
$\beta_{pl}=6.8$:
$$\vec{R}=(1,0,0), ~(2,0,0),~ (3,0,0),~ (4,0,0)$$
$$\vec{R}=(1,1,0),~ 2(1,1,0), ~(2,1,0),
~(1,1,1), ~(2,1,1), ~(2,2,1).$$
Summation over some possible paths from (0,0,0) to the final
destination points is done. This is a crude smearing.
The averages are projected back to $SU(3)$.

\subsection{Anisotropic lattice}

The tadpole-improved tree level action of Ref.\ \cite{Morn97} is used for
anisotropic lattices. The action has the form:
\begin{eqnarray}
S = \beta{\{\frac{5}{3}\frac{\Omega_{sp}}{\xi u_{s}^4}+
\frac{4}{3} \frac{\xi\Omega_{tp}}{u_{s}^2u_{t}^2}-
\frac{1}{12}\frac{\Omega_{sr}}{\xi u_{s}^6}-
\frac{1}{12}\frac{\xi\Omega_{str}}{u_{s}^4u_{s}^2 }}\},
\label{anisot}
\end{eqnarray}
where $\beta=6/g^2$, $g$ is the QCD coupling, and $\xi$ is the aspect ratio
($\xi=a_{s}/a_{t}$ at tree level in perturbation theory). $\Omega_{sp}$
and $\Omega_{tp}$ include the sum over spatial and temporal plaquettes;
$\Omega_{sr}$ and $\Omega_{str}$ include the sum over $2\times1$ spatial
rectangular and short temporal rectangular (one temporal and two spatial
links), respectively. For $a_{t} \ll a_{s}$ the discretization error of
this action is $O(a_{s}^4,a_{t}^2,a_{t}a_{s}^2)$. The coefficients are
determined at tadpole-improved tree level \cite{Lepa93}. The spatial
mean link, $u_{s}$, is given by:
\begin{equation}
\langle\frac{1}{3}ReTrP_{ss'}\rangle ^\frac{1}{4},
\end{equation}
where $P_{ss'}$ denotes the spatial plaquette. In general the temporal
link $u_{t}$ can be determined from:
\begin{equation}
u_{t} = \frac{\sqrt { \frac{1}{3} \langle Re Tr P_{st}\rangle}}{u_{s}},
\label{ut}
\end{equation}
where $P_{st}$ is the spatial-temporal plaquette. When
$a_{t} \ll a_{s}$, $u_{t}$, the temporal mean link can be fixed to
$u_{t}=1$, since its value in perturbation theory differs by unity by
$O(\frac{a_{t}^2}{a_{s}^2})$.

Measurements are done on four lattices $10^3\times24$,
$8^3\times24$, $18^3\times24$, and $16^3\times24$ at $\beta$ equal to
$1.7,~ 2.4,~ 2.4$ and $3.1$ with aspect ratios of $5,3,3,$ and $1.5$,
respectively. The temporal lattice spacings are kept approximately the same.
For the last lattice, $16^3\times24$ at $\beta=3.1$, the aspect ratio
is $1.5$, which means the spatial lattice spacing is only $1.5$ times 
bigger than the temporal lattice spacing. Therefore $u_{t}$ is 
calculated directly from Eq.\ (\ref{ut}).

$u_{s}$ and $u_{t}$ are spatial and temporal renormalization factors
introduced by mean field theory (``tadpole improvement''). Mean field
theory allows one to sum the effects of the largest higher order
contributions and thereby significantly improve the reliability
of perturbation theory.

At finite coupling, the anisotropy $\frac{a_{s}}{a_{t}}$ is renormalized
away from its input value $\xi$. Measurements of this renormalization can
be made by using the static quark potential extracted from correlations
along the different spatial and temporal axes of the lattice \cite{Morn97}.
The renormalization can be as large as $30\%$ without doing mean-link
improvement. With mean-link improvement, the difference is found to be
typically a few percent. In this work, I ignore the small radiative
corrections to the aspect ratio $\xi=\frac{a_{s}}{a_{t}}$. This includes
additional small lattice spacing errors which however vanish as
$a_{s} \rightarrow 0$.

\subsection{Smearing}

Smearing on spatial links is performed to
minimize the excited state contaminations in the correlation functions.
In other words, by smearing one can produce an interpolating field which
has greater overlap with the ground state and smaller overlap with
excited states. In the smearing procedure
each link is replaced by itself plus a sum of its four neighboring spatial
staples times a smearing factor $\lambda$ \cite{Alba87} (APE smearing):
\begin{eqnarray}
U_{j}(x) = P_{su(3)}\{ U_{j}(x)+
 \lambda \sum_{\pm k \neq j} U_{k}(x)
U_{j}(x+\hat{k})U^{\dagger}_{k} (x+\hat{j}) \}.
\label{smear}
\end{eqnarray}
$\lambda$ is the smearing factor and P indicates the projection back to
SU(3). Projection back to SU(3) after smearing or after averaging over
different paths in Wilson loops is necessary, in order to use 
Eqs.\ (\ref{r6}) to (\ref{r27}), in which the $U's$ must be SU(3) matrices.
For the same reason thermal averaging is not possible, since again it 
would take links out of SU(3). (Thermal averaging, which is normally 
useful to increase statistics, is the replacement of a time-like link 
by its average with fixed neighbors.)
The smearing factor, $\lambda$, is kept the same for all lattice
distances but the smearing levels (number of smearings) are chosen
separately for each distance from some low statistics runs.
With trial and error, one can optimize the smearing factor and smearing 
level. Here ``optimize'' means to find a good plateau
with good statistics at the smallest possible value of $T$.
In general for a given lattice, the
smearing level and the smearing factor are smaller for smaller $R$.
For large distance potentials, the configuration of glue appears to
spread out more, and more smearing is needed.
As a general rule, the finer the lattice spacing, the larger the smearing
level needed, so that the smearing extends out to a comparable physical
distance.

Projection back to $SU(3)$ is done by maximizing $tr(UV^{\dagger})$,
where $U$ is the normalized smeared link (in $SU(3)$) and $V$ is the
unnormalized smeared link. $V$ does not belong to $SU(3)$ since it is
obtained by summing the $SU(3)$ links (Eq.\ \ref{smear}). $U=\frac{V}{det(V)}$
minimizes $tr(UV^{\dagger})$ in $SU(2)$, because the sum of $SU(2)$
matrices is proportional to an $SU(2)$ matrix. To do this in $SU(3)$,
one loops over several embeddings of the $SU(2)$ subgroups in the
$SU(3)$ matrix; $U$ converges iteratively to the right answer.

\subsection{Fitting codes for analyzing the Wilson loops}

To take the auto-correlation length into account, the Wilson loops are 
grouped into 
blocks. The appropriate block size is determined by $j_{0}$, demanding
that the plaquette auto-correlation function fall near zero after
$j_{0}$ measurements. The error is stable under blocking with size
$j>j_{0}$. Using this method, the appropriate
block size has been chosen for each lattice simulation.

\begin{figure}
\epsfxsize=1. \hsize
\epsffile{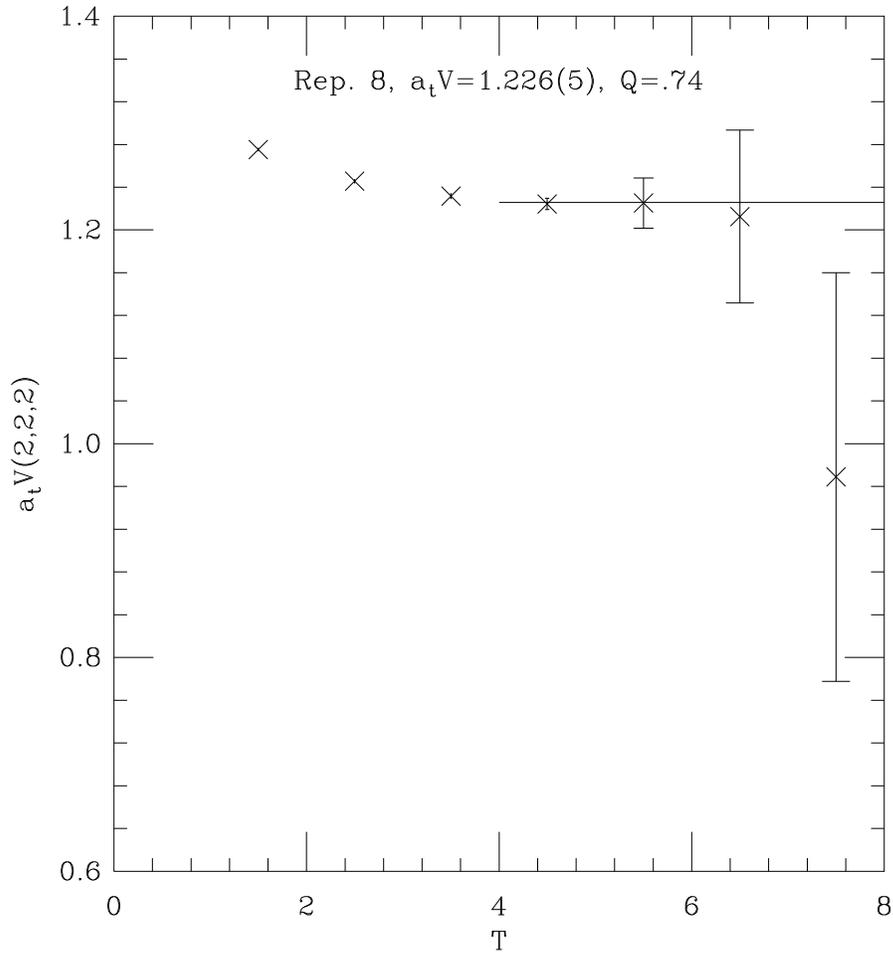}
\caption{$V(2,2,2)$ versus $T$ for $8^3\times24$ lattice and $\beta=2.4$.
The fit range is shown by the solid line.}
\label{gvsq12_8_2.4}
\end{figure}

Two fitting codes are used for computing the potentials and the string
tensions. The first code fits the Wilson loops of each fixed $R$ to the
exponential form:
$$ W \simeq \exp^{-V(R)T}$$
and finds the static quark potential $V(R)$ at each lattice distance $R$.
Fig.~\ref {gvsq12_8_2.4} shows a typical plot, obtained by the first 
fit, of the potential
between two static sources in representation 8 (adjoint) for the lattice
distance $\vec{R}=(2,2,2)$. The measurements in this figure have been 
obtained on an $8^3\times24$ lattice. The fit range, which is chosen 
individually for
each $R$, shows the plateau region. The errors of the fitting parameters 
are found by the jackknife method. $Q$, the confidence level for each 
fit, is calculated by measuring the covariance matrix evaluated by 
jackknife.

The potentials obtained by the first fit are then used to fit the
potential to the form:
\begin{equation}
V(R) \simeq -A/R + KR +C.
\label{VR}
\end{equation}
The results of the first fits are computed for all jackknife subensembles,
and therefore the jackknife errors of the second fit parameters can 
be computed.
The string tension, Coulombic coefficient, and the constant in lattice
units are the result of the second fit.

To have a better estimate of the covariance matrix, one can
try to eliminate its unreliable eigenvalues and eigenvectors
\cite{Harm67}. Suppose we have $N$
independent blocks of Wilson loops for a fixed $R$.
The average of Wilson loops at time $T_{i}$ is called $\bar{w}_{i}$:
\begin{equation}
\bar{w}_{i}=\frac{1}{N} \sum_{k} w_{ik},
\end{equation}
where $w_{ik}$ is the result for the $k$th block. Then the covariance
matrix is defined as:
\begin{equation}
C_{ij}= \frac{1}{N-1} (\frac{1}{N} \sum_{k=1}^N (w_{ik}-\bar{w_{i}})
        (w_{jk}-\bar{w_{j}}) ).
\end{equation}
It is then convenient to define the correlation matrix:
\begin{equation}
P_{ij} = \frac {C_{ij}} {\sqrt{C_{ii}C_{jj}}}.
\end{equation}
The diagonal elements of $P$ are equal to 1 and all the others are
less than 1.
Call the eigenvalues of the correlation matrix $\rho_{a}$.
Then $\sum_{a}\rho_{a}=n$ for the $n\times n$ correlation matrix.
The average eigenvalue is thus 1. The small eigenvalues $\rho_{a}\ll 1$
(or even $\rho_{a}< 1$) are often poorly determined with a finite
data set. These may give a false estimation of the covariance
matrix. One can make a new correlation matrix out of only those
eigenvectors whose eigenvalues are greater than some cutoff:
\begin{equation}
P' = B + D,
\end{equation}
where
\begin{equation}
B=\sum_{a}' \rho_{a} | \psi_{a} \rangle \langle \psi_{a}| ,
\label{B-eq}
\end{equation}
and
\begin{equation}
D_{ij}= \delta_{ij} (1-B_{ii}).
\end{equation}
The prime on the sum in Eq.\ (\ref{B-eq}) indicates that the sum is over
a subset of the eigenvectors: those whose eigenvalues are greater than
the cutoff.
Now $(C_{ij}')^{-1}$, which is the inverse of the modified covariance 
matrix, is given by:
\begin{equation}
(C_{ij}')^{-1}= \frac {(P_{ij}')^{-1}} {\sqrt{C_{ii}C_{jj}}},
\end{equation}
and
\begin{equation}
C_{ij}'=P_{ij}'\sqrt{C_{ii}C_{jj}}.
\end{equation}
If the cutoff is taken greater than the maximum
eigenvalue, then $C'$ would be a
diagonal matrix, with the diagonal elements equal to the standard errors,
and the fit would be an uncorrelated fit. This method thus interpolates 
between correlated and uncorrelated fits. In general, I keep the complete
covariance matrix in computing $Q$. However, in some cases the small
eigenvalues of the covariance matrix are poorly determined (due to limited
statistics), and I am forced to drop these eigenvalues and the
corresponding eigenvectors.

\subsection{Setting the scale}

I use the hadronic scale $r_{0}$, determined from
the force between static quarks at intermediate distance
\cite{Somm94}, to set the lattice scale. $r_{0}$ is defined by:
\begin{equation}
[r^2 \frac{dV(\vec{r})}{dr}]_{r=r_{0}}=1.65,
\label{r0}
\end{equation}
where $V(\vec{r})$ is the static quark potential in the fundamental
representation. The definition of Eq.\ (\ref{r0}) gives
$r_{0}\simeq 0.5$ fm in a phenomenological potential model. 
In this work I use $r_{0}^{-1}=410\pm 20$ MeV
determined by Morningstar \cite{Morn97}. 
This value is 
an average of $r_{0}$ from various quenched lattice simulations 
in which the quantities such as the mass of the $\rho$ or the $1P$-$1S$ 
splitting in heavy quarkonia are used to set the lattice spacing.
To fix $a_{s}$, the spatial lattice space, I thus use (see Eqs.\
(\ref{VR}) and (\ref{r0})):
\begin{equation}
\frac{r_{0}}{a_{s}} = \sqrt{ \frac{1.65-A}{Ka_{s}^2} }.
\end{equation}
For the anisotropic action I measure $Ka_{s}a_{t}$ from the fits. To
compute $Ka_{s}^2$, I have to know the aspect ratio
$\xi=\frac{a_{s}}{a_{t}}$. I use the
input value of $\xi$, since the difference between
the input value and what one computes from the lattice simulation
vanishes in the continuum limit and is
expected to be small at these lattice spacings \cite{Morn97}.
The scaling of the fundamental string tension is good
evidence that this expectation is correct.

\begin{figure}[p]
\epsfxsize=1. \hsize
\epsffile{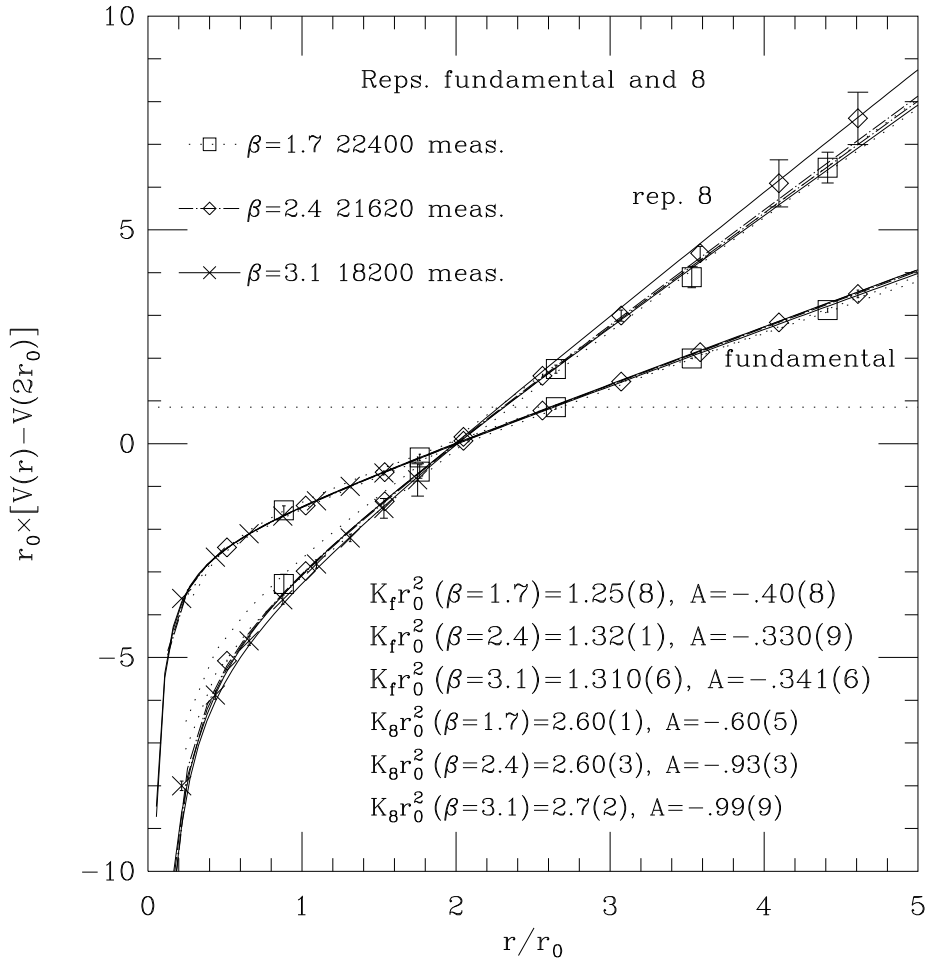}
\caption{The static quark potential $V(R)$ in terms of hadronic scale
$r_{0}$ for representations fundamental and 8. The double line
shows the central value of the fit $\pm$ error.
The horizontal dotted line shows the potential energy of glue-lumps.
Good scaling behavior is observed.
The weighted average string tensions for representations fundamental and 8 are 
$Kr_{0}^2$=1.324(4)(51) and $Kr_{0}^2$=2.602(9)(119), respectively.}
\label{scaled_comb}
\end{figure}

Using the scaled potentials and lattice distances in terms of the hadronic
scale $r_{0}$, it is possible to show the results of different lattice
simulations in one plot. For example see Fig.~\ref{scaled_comb}, where
$r_{0}[V(r)-V(2r_{0})]$ for the fundamental representation is plotted
versus $\frac{r}{r_{0}}$. All other physical quantities can be
computed in terms of the scale $r_{0}$ and can then be expressed
in relevant physical
units using the value of $r_{0}$ from other lattice simulations.
The scaled potentials for the higher representations can be plotted by using
the lattice spacing in terms of the hadronic scale of the fundamental
representation (see Fig.~\ref{scaled_comb} for representation 8).

\subsection{Computers and simulation codes}

In the first stage of this work, I have used code I have written in the
computer language $C$ for simulations on isotropic lattices. Later, to 
decrease the running time of the codes, I have used the MILC 
Collaboration software as a platform \cite{MILC}. The results obtained by
my previous codes on the isotropic lattices are in agreement with the 
results of the MILC platform codes.

The simulations have been done on a Dec Alpha and on Origin 2000's
supercomputers (single node jobs).

\section{Results and discussion}

\subsection{Isotropic Lattice}

Using the improved action, Eq.\ (\ref{isotropic}), the static quark
potentials for the fundamental representation have been found on an $8^4$ 
lattice at plaquette coupling constant, $\beta_{pl}$, equal to $6.8$. 
Other input parameters such as the rectangle and parallelogram coupling 
constants, have been computed from Eqs.\ (\ref{beta-rt}) to 
(\ref{alpha-s}) ($\beta_{rt}=-0.562$, $\beta_{pg}=-0.0844$).

\begin{figure}[p]
\epsfxsize=1. \hsize
\epsffile{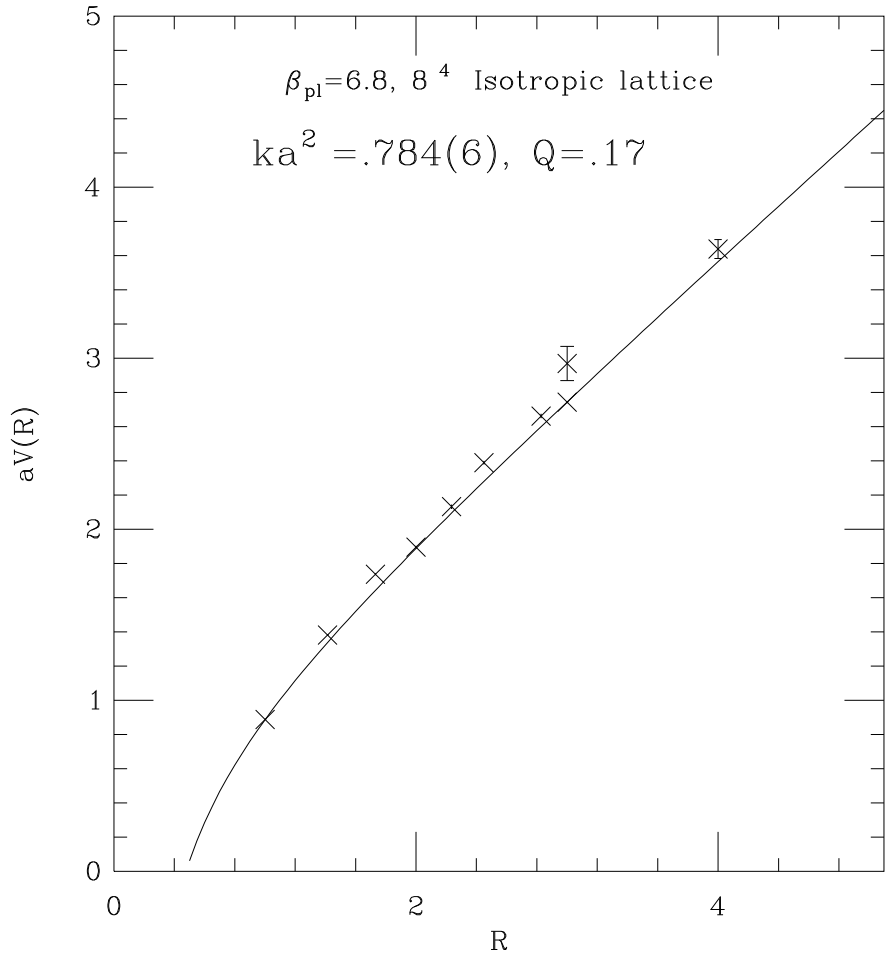}
\caption{Potential versus $R$ for an $8^4$ lattice at $\beta_{Pl}=6.8$.
$Ka^2$ shows the string tension times the square of the lattice spacing. $A$
indicates the Coulombic coefficient. The off-axis points deviate from the
fit by about $3-8\%$. Only on-axis points are used in the fit.}
\label{lrf_b=6.8}
\end{figure}

Fig.~\ref{lrf_b=6.8} represents the result of fitting the potentials
to the form of Eq.\ (\ref{VR}). The violation of rotational invariance
indicated by the deviation of
the off-axis potentials from the fit is about $3-8\%$, while with the same
lattice spacing, this deviation is as much as $35\%$ with Wilson action.
This deviation is due to finite lattice spacing errors. The
dimensionless string tension given by this fit is equal to 0.784(6).
By comparing this value with the phenomenological value of the string
tension $(420$MeV$)^2$, I find a lattice spacing of
approximately 0.42 fm.

\subsection{Anisotropic Lattice}

The simulations are performed for three coupling constants based on the
approximate desired spatial lattice spacing ratios 4:2:1. The aspect ratios have been chosen
such that the temporal spacings are approximately the same in all lattices. Two lattice
sizes for $\beta=2.4$, one for $\beta=1.7$ and one for $\beta=3.1$ are
studied. The renormalization factors for spatial
and temporal links, $u_{s}$ and $u_{t}$, have been computed for each
$\beta$. Table \ref{L-param} shows the input parameters used in the
simulations. The number of configurations represents the number
of measured Wilson loops for each $R$ and $T$. 
The last column of Table \ref {L-param} shows the
number of blocks for each calculation.

For all lattices except the $8^3\times24$ lattice, only on-axis potentials
have been measured. On $8^3\times24$, the potentials have been calculated
for the following distances:
$$\vec{R}=(1,0,0), ~(2,0,0),~ (3,0,0),~ (4,0,0)$$
$$\vec{R}=(1,1,0),~ 2(1,1,0), ~(2,1,0),~ (1,1,1),
~2(1,1,1), ~(2,1,1), ~(2,2,1).$$

In each lattice updating, three over-relaxation steps and one
Metropolis step are performed. Each Metropolis step includes $15$ hits on $SU(2)$
subgroups. Lattice measurements are done every three updates for
$\beta=1.7$ and $\beta=2.4$, and every four updates for $\beta=3.1$.

The potential at distance $R$ is determined from the asymptotic behavior
of Wilson loops, $W(R,T)$:
$$V(R)\simeq \lim_{T\rightarrow{+\infty}} \ln(\frac{W(R,T)}{W(R,T+1)})$$
Wilson loops have been fitted to the exponential form of
the above equation, and $V(R)$ times the temporal lattice spacing has
been found.

To extract the potential from the Wilson loops, one has to measure the
Wilson loops for large values of $R$ and $T$. 
To minimize the excited state contribution and to find the
static potential, one has to measure the Wilson loops for large $T$'s.
With the current statistics of this work (Table \ref{L-param}), it is
possible to go to large enough $T$ values to measure the Wilson loops
in lower representations and for small and even intermediate $R$ in
higher representations. For large $R$ values --- especially for higher 
representations --- the Wilson loops get too small for large $T$, and 
the error due to statistical fluctuations makes the measurements 
meaningless. Because of the lack of data for large values of $T$, I have 
therefore had to calculate the potentials using small $T$'s. However, 
I have been able to estimate the systematic error by changing the fit 
range or by comparing with $V$ of smaller $R$'s.
In most cases, it is still possible to get a good confidence level, $Q$,
after changing the fit range by one or two units in Tmin. The comparison
of the results for different values of Tmin gives an estimate of the
systematic error from excited states. However, for large values of $R$,
the Tmin's are usually very small. In these cases the systematic
error due to moving the plateau is taken from $V$ at smaller $R$.
I take the error on the potential as the sum in quadrature of statistical
and systematic errors. The systematic error is due to change of the fit
range.

\begin{figure}[p]
\epsfxsize=1. \hsize
\epsffile{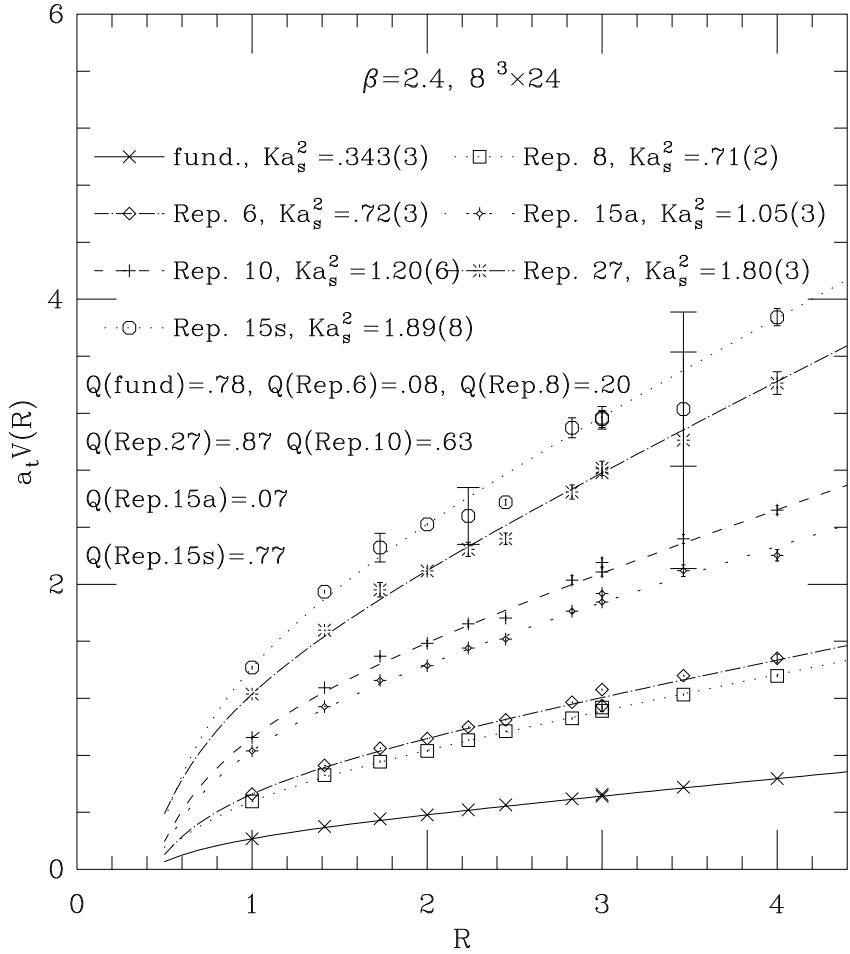}
\caption{A typical plot of $a_{t}V(R)$ versus $R$ for various
representations $8^3\times24$ lattice at $\beta=2.4$.
$Ka_{s}^2$ shows the string tension times the spatial lattice spacing
square. The error bars on the points are the sum in
quadrature of statistical and systematic errors. The systematic errors
are due to a change of the fit range of $V$ versus $T$.}
\label{rn_ref_2.4a}
\end{figure}

\begin{figure}[p]
\epsfxsize=1. \hsize
\epsffile{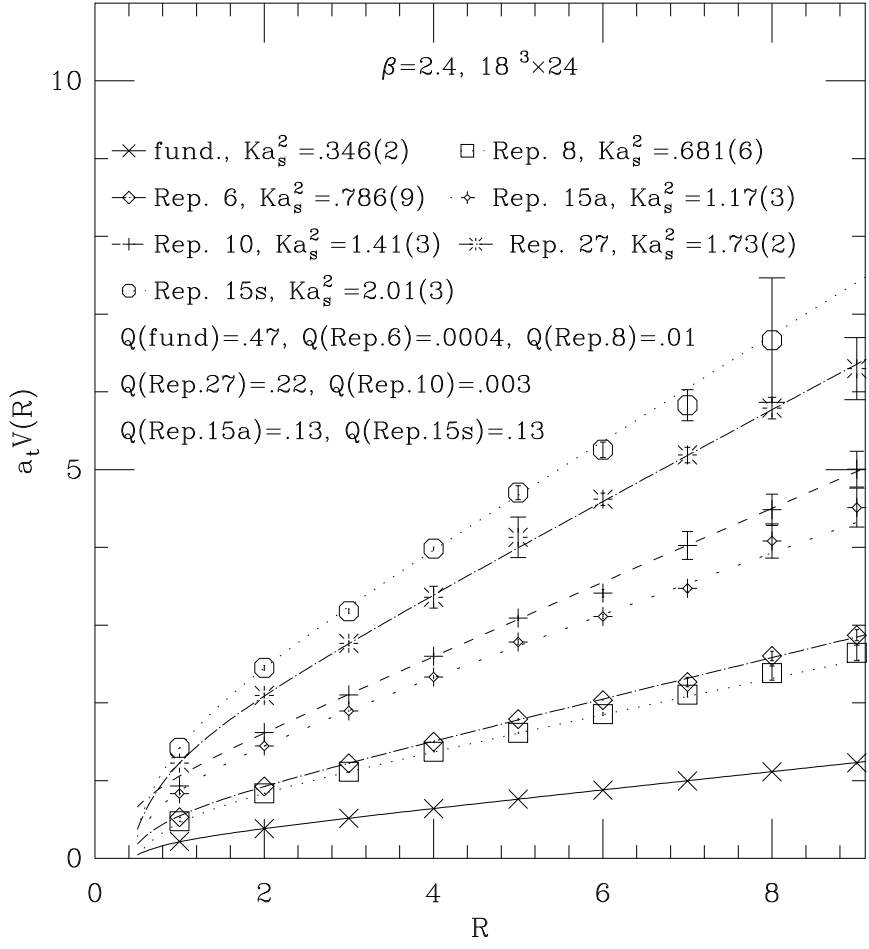}
\caption{Same as figure \ref{rn_ref_2.4a} but for the $18^3\times24$ lattice
at $\beta=2.4$.}
\label{rn_ref_2.4}
\end{figure}

The potentials obtained by fitting the Wilson loops to an
exponential form are then used to fit $V(R)$ versus $R$. 
A Gaussian distribution of the systematic error due to the
change of the fit range of the fits $V$ versus $T$ is added to the
potentials.
Then the modified data sets (potentials) from the first fits
are fitted to a Coulombic plus a linear term of Eq.\ (\ref{VR}).
As a result of this correlated fit, the
coefficient of the Coulombic term, $A$, the string tension times the
square of spatial lattice spacing, $Ka_{s}^2$, and the constant term
are found. The string tension in terms of
the hadronic scale $r_{0}$ is measured as well.
Fig.~\ref{rn_ref_2.4a} shows $a_{t}V$ versus $R$ (in units of $a_{s}$)
for fundamental, 6, 8, 15-antisymmetric, 10, 27 and 15-symmetric
dimensional representations for
$8^3\times24$ lattice
at $\beta=2.4$. Only on-axis points are used in the fits. Off-axis
potentials deviate from the fit about $2\%$-$8\%$; these are evidence
for rotational non-invariance -- a systematic error due to finite $a_{s}$.
Fig.\ \ref {rn_ref_2.4} shows the plot of $a_{t}V$ of
various representations versus lattice distance $R$ for
$\beta=2.4$ ($18^3\times24$ lattice). The confidence levels of the fits
are indicated by $Q$'s. The error bars on the points are the sum in
quadrature of statistical and systematic errors due to change of the fit range in $T$.
The slopes of the potentials are qualitatively in agreement with Casimir
scaling.
The fundamental-representation results from the fits of $V$ versus $R$
for different lattices are shown in Table \ref{sec_fund}. In this Table
the spatial lattice spacing $a_{s}$ in terms of $r_{0}$ is given for
different lattices. The temporal lattice spacing, $a_{t}$, can be
found by aspect ratio, $\xi=a_{s}/a_{t}$. 

\begin{figure}[p]
\epsfxsize=1. \hsize
\epsffile{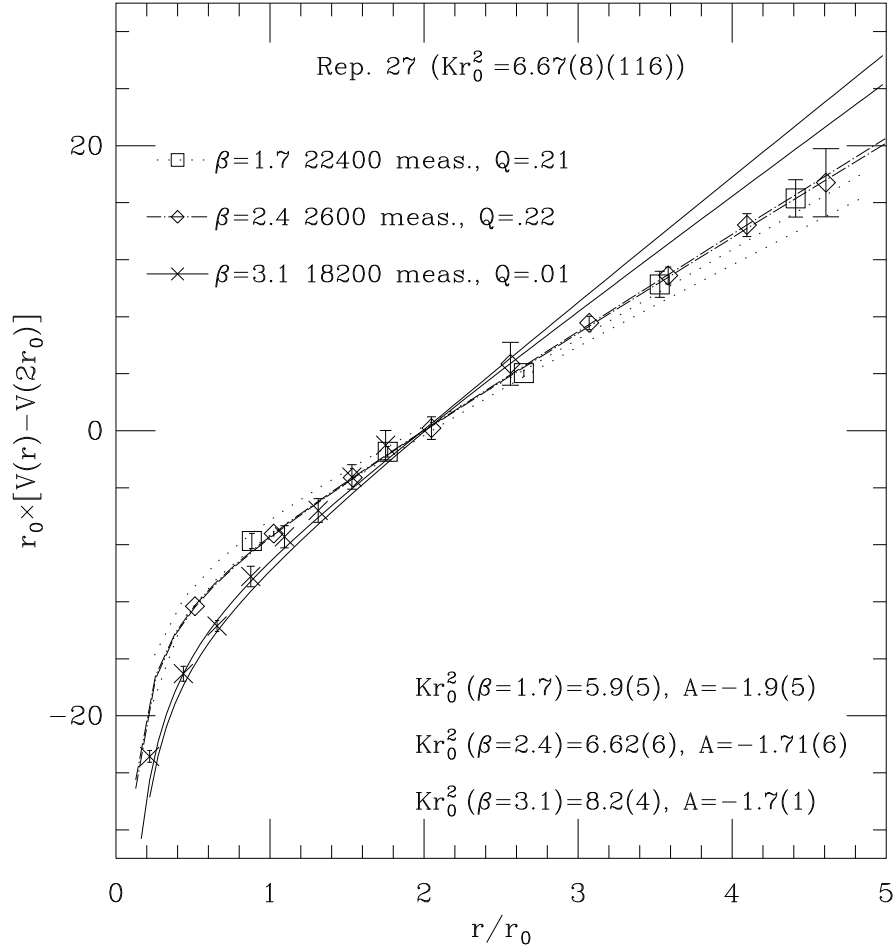}
\caption{Same as figure \ref{scaled_comb} but for representation 27. The
final result is $Kr_{0}^2$=6.67(8)(116).}
\label{scaled_27}
\end{figure}

Figs.~\ref{scaled_comb} and
\ref{scaled_27} show the static quark potential
of some representations in terms of the hadronic scale for $\beta=1.7$, 
$\beta=2.4$ and $\beta=3.1$. The dimensionless string tension, $Kr_{0}^2$,
for each $\beta$ and the best estimate for each representation are 
presented. $A$ shows the dimensionless Coulombic coefficient.
The best estimate for the string tension of each representation 
is obtained by the weighted average of the four lattice measurements. The 
errors in the string tensions are the statistical error (from the 
weighted average) and the systematic error of discretization 
(determined by the standard deviation of the results over
the 3 couplings), and the error on $r_{0}$, respectively.
Table \ref {best-kr0} gives $Kr_{0}^2$ for different $\beta$'s and the 
best estimate of the dimensionless string tension.
In Table \ref{k0-ener} the ratio of the string tension of each 
representation to the fundamental one is given. 
There is a rough agreement between these ratios and the Casimir number
ratios of higher representations to the fundamental one as predicted 
\cite{Ambj84}. The fundamental representation
string tension, 0.222(1)(8)(21) GeV, is in reasonable agreement 
with the phenomenological value of 0.18(2) GeV. (Note that the value used
here for $r_{0}$, $r_{0}^{-1}=410\pm 20$ MeV, is determined without
reference to the string tension \cite{Morn97}.) Based on the best estimate of the string tension
given by Table \ref {k0-ener}, $K_{8}/K_{f}=1.97(1)(12)$. This ratio is
in agreement with 2.2(4) reported by Campbell {\it et al.}
\cite{Camp86}.

The Coulombic coefficients ($A$), the best estimate of the $A$'s 
found by weighted average, and the ratios of the average $A$ of each 
representation to the the fundamental one are given in Table \ref{best-A}. 
Comparing the ratios with Casimir ratios, one can see some qualitative
agreement, although there are some large disagreements for a few of the
representations. The value of the Coulombic coefficient, $A$, is quite
sensitive to the lower bound of the fit range. That is why it shows some
differences among different lattices. The weighted average of the 
fundamental Coulombic coefficient, $A$, from Table \ref{sec_fund}
is equal to $.329(4)(49)$. This is about 
$2 \sigma$ off from $\frac{\pi}{12}$ of the standard infrared 
parametrization form of the static quark potential: 
$$V(R) \simeq -\pi/12 + KR +C.$$

For the fundamental representation, good scaling behavior (results 
independent of lattice spacing) is observed. The scaling gets worse for 
higher representations, roughly in proportion to the string tension. 
It seems that the higher representations are
more sensitive to the lattice spacing although it is difficult to make a
definitive statement because the data get progressively worse as the
dimension of the representation increases.
With the current data, the string tensions of the higher representations
at intermediate lattice spacing ($\beta=2.4$) are actually closer to
the result at the coarsest lattice spacing ($\beta=1.7$) than they are to the
result of the finest lattice spacing ($\beta=3.1$). Scaling errors are the 
most important systematic error in this study.

Screening or a change of the slope should start to show up when
the potential energy of the static quarks becomes equal to or larger than
twice the glue-lump mass. As a result, a pair of gluons pops out of vacuum
which interact with the static sources and cause screening or a changing
of the potential slope, depending on the triality of the representation. The
horizontal dotted line in Fig.~\ref {scaled_comb} shows twice the
glue-lump potential. The static potential energy of the two adjoint sources
becomes equal to $2\times M_{glueball}$ at about 1.2 fm \cite{Mich98}. This
energy is enough to pop out two octets.
In the current work, no change of the slope of the linear part of the potential
for representations 6, 15-antisymmetric and
15-symmetric is observed yet (they can in principle be converted into the
fundamental representation by dynamical gluons).
No screening for representations 8, 10, 27 is seen either.
(These representations
have zero triality and can in principle be screened by dynamic gluons.)
This is not surprising, given the well known fact that the Wilson loops
do not couple well to screened states. Presumably one would need
additional sources which couple better to separated screened states
(see, e.g.,) \cite{Step99} and \cite{Phil99}) to
see the screening at reasonable values of $R$. 
Wilson loops couple primarily to string-like, confined states.

Using the hadronic scale $r_{0}$
given by $r_{0}^{-1}=410\pm 20$ MeV \cite{Morn97}, one can
calculate the lattice spacings in energy units.
The spatial lattice spacings are then 0.43(1)(2) fm, 0.2482(6)(101) fm, 
0.2509(5)(102) fm, and 0.1072(5)(44) fm for lattices
$10^3\times24$, $8^3\times24$, $18^3\times24$, and $16^3\times24$ at
$\beta$ equal to $1.7,~ 2.4,~ 2.4$, and $3.1$, respectively. The first
error of each lattice spacing is the error from this work and the second 
error is due to the hadronic scale uncertainty.

The finite volume effect is studied by measuring the Wilson loops on
the two lattice sizes $8^3\times24$ and $18^3\times24$ for $\beta=2.4$.
The potentials are generally in agreement; in a very few cases
a slight difference is observed (not more than $6\%$). The scaled
string tensions, $Kr_{0}^2$, of Table \ref{best-kr0} are also in good
agreement for the two lattices.
A difference of at most $2\sigma$ is observed for some
representations. This can be explained as a result of a different fit
range for the smaller lattice. In other words, for the bigger lattice,
$18^3\times24$, larger distances have a more significant role in
determining the string tension. In
contrast, for the smaller lattice, $8^3\times24$, the fits are based on
only four points. Since the physical volume at $\beta=1.7$ and $\beta=3.1$
are comparable to or larger than that at $8^3\times24$, $\beta=2.4$,
I conclude that the lattice volumes are big enough
and the finite volume errors are much smaller than the dominant systematic
error, which is due to discretization. Figs.~\ref{rn_ref_2.4a}
and \ref{rn_ref_2.4} show the potentials and the string tensions for
various representations for the two lattice sizes.

Another source of the systematic error is due to the finite lattice
spacing in the time direction. The action used for the anisotropic 
lattice (Eq.\ (\ref{anisot})) has
$O(a_{s}^4, a_{t}^2, \alpha_{s}a_{s}^2)$ discretization errors and is
intended for use with $a_{t}\ll a_{s}$.
As I take $a_{s}$ smaller in this work, I keep $a_{t}$ fixed. At
$\beta=3.1$, the aspect ratio, $\xi=\frac{a_{s}}{a_{t}}$ is only $1.5$,
so the $O(a_{t}^2)$ errors may no longer be negligible compared to the
$O(a_{s}^4)$ errors, although they are expected to be small on an
absolute scale.

\section{Conclusion}

The string tensions in various representations are basic properties of
the pure non-Abelian gauge theory and therefore should be predicted
in any theory of confinement for any representation.
In this work, the static potentials for the
fundamental and some higher representations have been studied by
numerical lattice calculations. The results provide a test
of confinement models.

By lattice calculations, the fundamental static potentials and
the string tension have been computed by measuring the Wilson loops on an
$8^4$ isotropic lattice at $\beta_{pl}=6.8$. The one-loop,
tadpole-improved QCD gauge action has been used. Both on- and off-axis
potentials have been measured, and the effect of the coarse lattice has
been discussed. It has been found that the violation of rotational
invariance indicated by the deviation of the off-axis potentials from
the fit is only about $3-8\%$. This makes the improved action a useful
tool in lattice calculations.

The fundamental and higher representation potentials and the string
tensions have been computed on four anisotropic lattices using a
tadpole-improved tree level action. To extrapolate to the continuum and
to study the volume effect, three lattice spacings with the approximate
ratio of 1:2:4 on four lattice sizes have been studied. Lattice spacings
range from 0.11 fm to 0.43 fm and the lattice sizes are
$10^3\times24$, $8^3\times24$, $18^3\times24$, and $16^3\times24$ at
$\beta$ equal to $1.7,~ 2.4,~ 2.4$, and $3.1$, respectively.
Using the hadronic scale $r_{0}$, which is defined in terms of the 
potential between static quarks at an intermediate distance in the 
fundamental representation, the dimensionful string tensions have 
been calculated, and the scaling behavior has been studied. Good 
scaling behavior for the fundamental representation is observed. 
The scaling gets worse for higher representations, roughly in 
proportion to the string tension.
At small and intermediate distances, rough agreement with Casimir scaling 
is observed. 

\section{Acknowledgments}

I would like to thank Claude Bernard for his great support in 
this work. I wish to thank the MILC Collaboration and especially Robert 
Sugar and Steven Gottlieb for computing resources. Also I wish to thank 
J. Mandula, U. Heller and C. Morningstar for helpful discussions.

\clearpage

\begin{table}[p]
\setlength{\tabcolsep}{.25pc}
\caption{Input parameters for lattice calculations. The renormalization
factors for spatial and temporal links are given by $u_{s}$ and $u_{t}$,
respectively.}
\label{L-param}
\begin{center}
\begin{tabular}{lccccc}
\hline
Lattice  &{$\beta$} & $u_{s}^4$  & {$u_{t}$} & {$\#$ configs} & $\#$ blocks
  \\
\hline\\
$10^3\times 24$   & $ 1.7$  & $.295$ & $1.$    & $22400$      & $224$   \\
 \\
$8^3\times 24$    & $ 2.4$  & $.421$ & $1.$    & $7200-7900$  & $50$ \\ \\
$18^3\times 24$   & $ 2.4$  & $.421$ & $1.$    & $21620$      & $188$  \\ \\
$16^3\times 24$   & $ 3.1$  & $.575$  & $.945$ & $18200$      & $182$\\
\hline
\end{tabular}
\end{center}
\end{table}

\begin{table}[p]
\setlength{\tabcolsep}{.1pc}
\caption{Results of the fit of the fundamental potential to a Coulombic plus linear
form. $A$ shows the Coulombic coefficient and $Cr_{0}$ is the constant
term times the hadronic scale $r_{0}$. The fundamental string tension
and the spatial lattice spacing in terms of $r_{0}$ are given as
$Kr_{0}^2$ and $r_{0}/a_{s}$, respectively.}
\label{sec_fund}
\begin{center}
\begin{tabular}{lcccccc}
\hline
$\beta$  & Lattice & $\xi$   & $r_{0}/a_{s}$ &  $A$   & $Kr_{0}^2$ &  $Cr_{0}$    \\
\hline\\
$1.7$ & $10^3\times24$  & $5.0$  & $1.13(3)$ &  $-.40(8)$  & $1.25(8)$ & .42(7)       \\ \\
$2.4$ & $8^3\times24$   & $3.0$  & $1.974(5)$ &  $-.315(6)$  & $1.335(5)$ & 1.22(2)    \\ \\
$2.4$ & $18^3\times24$  & $3.0$  & $1.953(4)$ &  $-.330(9)$  & $1.32(1)$ &
 1.21(1)    \\ \\
$3.1$ & $16^3\times24$  & $1.5$  & $4.57(2)$ &  $-.341(6)$  & $1.310(6)$ &
 3.07(3)    \\ \\
\hline
\end{tabular}
\end{center}
\end{table}

\begin{table}[p]
\setlength{\tabcolsep}{1.pc}
\caption{String tensions in terms of $r_{0}$ for different coupling
constants, lattice sizes, and the best estimate.}
\label{best-kr0}
\begin{center}
\begin{tabular}{lccccc}
\hline
Rep.  & $Kr_{0}^2(\beta=1.7)$   & $Kr_{0}^2(\beta=2.4)$ &
$Kr_{0}^2(\beta=2.4)$ & $Kr_{0}^2(\beta=3.1)$ & Best estimate    \\
  & $10^3\times24$  & $8^3\times24$ & $18^3\times24$ & $16^3\times24$ & \\
\hline\\
3   & $1.25(8)$ & $1.335(5)$ & $1.32(1)$ &  $1.310(6)$  & $1.324(4)(51)$
      \\ \\
8   & $2.60(1)$ & $2.75(8)$   & $2.60(3)$  &  $2.7(2)$  & $2.602(9)(119)$
      \\ \\
6   & $2.9(2)$  & $2.8(2)$  & $3.00(3)$ &  $3.58(8)$  & $3.06(3)(40)$
   \\ \\
15a & $4.4(2)$  & $4.1(2)$   & $4.6(1)$ &  $5.7(3)$  & $4.57(8)(82)$
   \\ \\
10  & $4.9(3)$  & $4.7(2)$   & $5.4(2)$ &  $7.5(2)$  & $5.8(1)(16)$
  \\ \\
27  & $5.9(5)$  & $7.0(4)$   & $6.62(6)$  &  $8.2(4)$  & $6.67(8)(116)$
     \\ \\
15s & $7.1(5)$  & $7.4(4)$   & $7.6(2)$  &  $12.2(2)$  & $9.5(1)(31)$
    \\ \\
\hline
\end{tabular}
\end{center}
\end{table}

\narrowtext
\begin{table}[p]
\setlength{\tabcolsep}{.5pc}
\caption{Best estimate of string tensions in energy units from different
coupling constants. The ratio of string
tensions is proportional to the ratio of Casimir scaling of the last
column.}
\label{k0-ener}
\begin{center}
\begin{tabular}{lccc}
\hline
Rep.  & K (GeV) & $\frac{K_{r}}{K_{f}}$ & $\frac{C_{r}}{C_{f}}$\\
\hline
3   &    $0.222(1)(8)(21)$  &    -           &    -           \\ \\
8   &    $0.437(2)(20)(42)$ & $1.97(1)(12)$  & $2.25$  \\ \\
6   &    $0.514(5)(67)(49)$ & $2.32(3)(31)$  & $2.5$    \\ \\
15a &    $0.77(1)(15)(7)$   & $3.47(5)(69)$  & $4.0$      \\ \\
10  &    $0.97(2)(27)(9)$   & $4.37(9)(123)$ & $4.5$      \\ \\
27  &    $1.12(1)(20)(11)$  & $5.05(5)(92)$  & $6$       \\ \\
15s &    $1.60(2)(52)(15)$  & $7.2(1)(24)$   & $7$       \\
\hline
\end{tabular}
\end{center}
\end{table}

\widetext
\begin{table}[p]
\setlength{\tabcolsep}{.05pc}
\caption{Coulombic coefficients found by different lattice calculations
and the best estimate. Rough agreement with Casimir ratios is observed.}
\label{best-A}
\begin{center}
\begin{tabular}{lccccccc}
\hline
Rep.  & $A(\beta=1.7)$   & $A(\beta=2.4)$ &
$A(\beta=2.4)$ & $A(\beta=3.1)$ & best estimate & $\frac{A_{f}}{A_{r}}$ & 
$\frac{C_{r}}{C_{f}}$    \\
  & $10^3\times24$  & $8^3\times24$ & $18^3\times24$ & $16^3\times24$ & & & \\
\hline\\
3   & $-.40(8)$ & $-.315(6)$ & $-.330(9)$ &  $-.341(6)$  & $-.329(4)(43)$
    & - & -  \\ \\
8   & $-.60(5)$ & $-.75(6)$   & $-.93(3)$  &  $-.99(9)$  & $-.84(2)(18)$
    & $2.55(7)(64)$  & $2.25$  \\ \\
6   & $-.54(9)$  & $-.93(9)$  & $-.69(6)$ &  $-.74(5)$  & $-.72(3)(16)$
    & $2.2(1)(6)$ & $2.5$\\ \\
15a & $-.84(1)$  & $-1.5(1)$   & $-1.2(2)$ &  $-1.23(9)$  & $-.85(1)(48)$
    & $2.58(4)(150)$ & $4.0$\\ \\
10  & $-.50(2)$  & $-1.6(2)$   & $-.5(2)$ &  $-.7(1)$  & $-.52(2)(63)$
    & $1.58(6)(193)$ & $4.5$\\ \\
27  & $-1.9(5)$  & $-1.6(2)$   & $-1.71(6)$  &  $-2.5(2)$  & $-1.76(6)(45)$
    & $5.4(2)(16)$ & $6$ \\ \\
15s & $-1.6(4)$  & $-2.3(2)$   & $-2.1(2)$  &  $-1.95(6)$  & $-1.98(6)(30)$
    & $6.0(2)(12)$ & $7$  \\ \\
\hline
\end{tabular}
\end{center}
\end{table}
\end{document}